# Playing with words: Do people exploit loaded language to affect others' decisions for their own benefit?

Valerio Capraro[1], Andrea Vanzo[2], and Antonio Cabrales[3]

[1] Middlesex University London, United Kingdom
[2] Artificialy SA, Switzerland
[3] Universidad Carlos III de Madrid, Spain

Contact: VC, v.capraro@mdx.ac.uk; AV, andrea.vanzo@artificialy.com, AC, antonio.cabrales@uc3m.es

**Abstract**

We report on three pre-registered studies testing whether people in the position of describing a decision problem to decision-makers exploit this opportunity for their benefit, by choosing descriptions that may be potentially beneficial for themselves. In Study 1, recipients of an extreme dictator game (where dictators can either take the whole pie for themselves or give it entirely to the receiver) are asked to choose the instructions used to introduce the game to dictators, among six different instructions that are known from previous research to affect dictators' decisions. The results demonstrate that some dictator game recipients tend to choose instructions that make them more likely to receive a higher payoff. Study 2 shows that people who choose descriptions that make them more likely to receive a higher payoff indeed believe that they will receive a higher payoff. Study 3 shows that receivers are more likely than dictators to choose these descriptions. In sum, our work suggests that some people choose descriptions that are beneficial to themselves; we also found some evidence that deliberative thinking and young age are associated with this tendency.

**Introduction**

Recent work has shown that people's decisions in economic experiments involving prosocial behavior do not only depend on the monetary consequences that the available actions bring about, but they also depend on the specific words that are used to describe these actions, especially when they activate moral concerns. For example, Eriksson et al. (2017) found that rejection rates in the ultimatum game depend on the label used to describe the rejection option: people are less likely to reject low offers when the rejection option is labelled as "payoff reduction" compared to when it is labelled as "costly punishment"; moreover, this effect could be explained by the moral judgments associated with the available actions. Capraro and Rand (2018) found that decisions in a trade-off game (where decision-makers get to decide between an equal and an efficient allocation of money) are affected by the words used to describe the available actions[1]; again, this effect was explained by differences in moral judgments. A similar finding was reported by Capraro and Vanzo (2019) using an "extreme" dictator game. Specifically, they randomly assigned participants to one of six extreme dictator game frames. In all frames, the dictator was informed that there was a certain amount of money available for themselves and the recipient. The dictator could decide between two options: to allocate all the money to themselves, or to allocate all the money to the recipient. The six frames differed on the wording used to describe such options, which were meant to trigger moral concerns. In the *steal* frame, the self-regarding allocation explicitly presented the words "steal the money", whereas the other-regarding allocation used the words "don't steal the money". The *take*, the *demand*, the *give*, the *donate*, and the *boost* frames were analogue. Results showed that participants assigned to the *steal* frame were the most likely to implement the other-regarding allocation, whereas participants assigned to the *boost* frame were the least likely[2]; once again, this effect could be explained by the moral judgments associated with the available actions. Chang et al. (2019) also found that dictator game giving depends on the words used to describe the available actions: politically framed words interact with participants' political orientation to determine changes in norms and decisions. Finally, Mieth et al. (2021) found that the rate of cooperation in the prisoner's dilemma changes if the available options are described using non-neutral, morally loaded, words, as "I cooperate" and "I cheat".

---

[1] This result has been replicated several times in different experimental contexts (Tappin & Capraro, 2018; Huang et al. 2019; Capraro, Rodriguez-Lara, Ruiz-Martos, 2020; Capraro, Jordan, Tappin, 2020; Huang et al. 2020).

[2] Somewhat related is the literature comparing the dictator game in the *give* vs *take frame* (Swope et al. 2008; Dreber, Ellingsen, Johannesson, & Rand, 2013; Krupka & Weber, 2013; Grossman and Ecker, 2015; Halvorsen, 2015; Hauge, Brekke, Johansson, Johansson-Stenman, & Svedsäter, 2016; Goerg, Rand, & Walkowitz, 2019). In this stream of literature, the two frames do not only differ in the words used to describe the actions, but also in where the endowment is initially allocated: in the *give frame*, it is initially allocated to the dictator; in the *take frame*, it is initially allocated to the recipient, or split between the dictator and the recipient. These experimental works showed mixed results, providing no clear evidence that one frame gives rise to more prosociality than the other. Consistent with these mixed results, also Capraro and Vanzo (2019) reported no statistical difference between their *give* and *take* frames. The two approaches are related because it is likely that the words used to describe the available actions generate a continuous endowment effect (Reb & Connolly, 2007).

Now, if decision-makers' choices are affected by the words being used to describe the available actions, it is possible that people in the position of describing a decision problem to decision-makers exploit this opportunity for their benefit, by choosing descriptions that may potentially favor themselves. In this article we study this issue. We believe this to be an important question from the practical viewpoint, as, in several real life situations, people are indeed in the position of describing a decision problem to a decision maker. The same group of activists may be labelled as terrorists or freedom-fighters depending on the impact the provider of information wants to elicit from the receiver (Ganor, 2002). Similarly, a voting process can be labelled a referendum or an illegal act (Pascual Planchuelo, 2020).

In this article, we capture the core of this issue by taking the mirror image of the methodology introduced by Capraro and Vanzo (2019). Specifically, in Study 1 recipients of an extreme dictator game are asked to choose the frame used to introduce the game to dictators, among *steal*, *demand*, *take*, *donate*, *give*, *boost*, as above. If recipients do not exploit the opportunity to choose the frame to affect the dictators' decisions, each frame will be selected following the uniform distribution. Conversely, if instead some recipients exploit the opportunity for their benefit, then the frame that they believe to be most likely associated with the other-regarding allocation will be chosen more frequently. From Capraro and Vanzo (2019), we know that this frame is the *steal* frame. Therefore, assuming that a significant proportion of those recipients, who wants to exploit the chance to choose the frame for their benefit, have correct beliefs about which frame gives rise to the maximum rate of other-regarding behavior, we obtain our primary (and pre-registered) hypothesis:

*Hypothesis 1. Recipients are more likely to choose the steal frame than what one would expect from uniformly random choice.*

An important remark is due at this stage. As said above, Capraro and Vanzo (2019) also found that people assigned to the *boost* frame were the least likely to implement the other-regarding allocation. However, this does not allow us to logically conclude that, in Study 1, we will observe that recipients choose the *boost* frame less frequently than the other frames. An alternative is indeed possible, at least in principle. Assume that there are two classes of people: (i) those who choose the frame uniformly at random and (ii) those who choose the *steal* frame. In this case, we would observe all frames to be chosen equally often, apart from the *steal* frame, which would be chosen more often than the other frames. Therefore, the only hypothesis that we can deduce from Capraro and Vanzo (2019) is Hypothesis 1.

Coming back to this hypothesis, while we expect it to be confirmed in our experiment, we do not expect that all recipients choose the *steal* frame. It is likely that there are individual characteristics that are associated with the probability of choosing this frame. Which characteristics? Among the basic demographic variables, we may expect gender and age to play a role, because they are known to be associated with dictator game decisions: several meta-analyses have shown that women tend to be more altruistic than men (Engel, 2010; Brañas-Garza et al. 2018; Rand et al. 2016) and that older people tend to be more altruistic than younger people (Engel, 2010). Therefore, these demographic characteristics could be associated with the probability of choosing the *steal* frame.

Apart from demographic characteristics, we reasoned that other two individual factors might be associated with the probability of choosing the *steal* frame. One is deliberative thinking: understanding that the frame can be used to affect dictators' decisions might require a non-negligible amount of deliberation. The other one is strategic sophistication: understanding that the frame can be strategically used to affect dictators' decisions could require some amount of strategic sophistication. To measure deliberative thinking, we used the Cognitive Reflection Test (CRT; Frederick, 2005). To measure strategic sophistication, we used the 11-20 task (Arad & Rubinstein, 2012). We refer to the Method section for detailed descriptions of these tasks.

We conducted a first session of Study 1 to test Hypothesis 1 and to explore the effect of gender, age, deliberative thinking, and strategic sophistication. We pre-registered Hypothesis 1 and, as secondary analyses, we pre-registered that we would explore the effect of gender, age, deliberative thinking, and strategic sophistication (see https://aspredicted.org/ux4h3.pdf). Hypothesis 1 was immediately confirmed. Moreover, we found some evidence, although weak, of an effect of deliberative thinking, such that participants with higher CRT scores were more likely to use the *steal* frame. Therefore, we conducted another session (pre-registered at: https://aspredicted.org/ru8fz.pdf) with the goal of replicating Hypothesis 1 and, additionally, of testing the following pre-registered hypothesis:

*Hypothesis 2. Recipients with higher scores in the CRT are more likely to choose the steal frame.*

Hypothesis 1 was successfully replicated. As for Hypothesis 2, we did not find strong evidence that higher scores in the CRT are associated with the probability of choosing the *steal* frame.

Note that Hypothesis 1 does not imply that receivers choose the *steal* frame strategically in order to increase the likelihood to receive a higher payoff. It is possible that receivers are more likely to choose the *steal* frame because of some other, possibly non-strategic, reasons (e.g., because they like the *steal* frame). To rule out this possibility, we conducted two additional studies, providing two pieces of evidence in support of our interpretation that some receivers choose the *steal* frame strategically to increase the probability to receive a higher payoff. Specifically, in Study 2 we test the following pre-registered hypothesis:

*Hypothesis 3. Receivers who choose the steal frame are more likely than receivers who choose the boost frame to believe that they will get the higher payoff.*

In Study 3 we test the following pre-registered hypothesis:

*Hypothesis 4. Receivers are more likely than dictators to choose the steal frame.*

Both these hypotheses were confirmed by the results. Furthermore, in Studies 2-3, we found support also for Hypothesis 2.

## Study 1

**Method**

Participants, based in the USA at the time of the experiment, were recruited on Amazon Mechanical Turk (AMT; Arechar, Gächter, & Molleman, 2018; Paolacci & Chandler, 2014; Paolacci, Chandler, & Ipeirotis, 2010). After entering their TurkID and accepting the informed consent, they were told that they would be playing a game involving two participants, Participant A and Participant B, and an amount of money ($0.50). They were informed that there were two available options, Option 1 and Option 2. Option 1 allocates the whole amount to Participant A, Option 2 allocates the whole amount to Participant B. Then, they were informed that the role of Participant B is to decide the wording used to present Option 1 and Option 2 to Participant A, and that the role of Participant A is to decide between Option 1 and Option 2, as presented using the wording decided by Participant B. We clarified that Participant A will not be made aware that Participant B chose the descriptions of the options; Participant A will only play the game as described by Participant B, without ever knowing where these descriptions come from. Participants were then asked five comprehension questions to make sure that they understood the game. We automatically excluded participants that failed one or more comprehension questions. These participants received the participation fee ($0.50), but no additional bonus. Participants were informed of this exclusion criterion at the beginning of the survey. All participants who passed the comprehension questions were informed that they would be playing as Participant B (to compute their payoff, we matched them with participants in the experiment reported by Capraro and Vanzo (2019), which were treated as Participants A - see below for further details). Participants were reminded about their role and that of Participant B. Then, they were shown six possible descriptions of Option 1 and Option 2:

Option 1 = Participant A steals all the money from Participant B
Option 2 = Participant A does not steal all the money from Participant B

Option 1 = Participant A takes all the money from Participant B
Option 2 = Participant A does not take all the money from Participant B

Option 1 = Participant A demands all the money from Participant B
Option 2 = Participant A does not demand all the money from Participant B

Option 1 = Participant A does not give all the money to Participant B
Option 2 = Participant A gives all the money from Participant B

Option 1 = Participant A does not donate all the money to Participant B

Option 2 = Participant A donates all the money from Participant B

Option 1 = Participant A does not boost Participant B with all the money
Option 2 = Participant A boost Participant B with all the money

These descriptions were shown in random order.[3] Participants made their decision by ticking on their preferred pair of descriptions.

After making their decision, participants played the 11-20 game to measure strategic sophistication (Arad & Rubinstein, 2012). In this game, participants are randomly matched to play a game against another participant. Each participant requests an amount of money between 11 and 20 cents. Each participant receives the amount they request, plus an additional 20 cents if they ask for exactly one cent less than the other participant. This game was incentivized. The strategic structure of this game makes it particularly suitable to measure participants' strategic sophistication using Level-k theory (Stahl & Wilson, 1994). Choosing 20 is intuitively appealing and is the instinctive choice when choosing a sum of money between 11 cents and 20 cents; it also guarantees the highest certain payoff, while avoiding strategic thinking. For these reasons, choosing 20 can be considered a Level-0 strategy (Arad & Rubinstein, 2012). Now note that choosing 19 is the best response to choosing 20; moreover, choosing 19 is also the best response to the uniformly random choice. Therefore, choosing 19 can be considered as a Level-1 strategy. Similarly, choosing 18 can be considered as a Level-2 strategy, and so on (Arad & Rubinstein, 2012).

After the 11-20 game, participants took a 3-item Cognitive Reflection Test (CRT; Frederick, 2005). Each item consists of a question characterized by the fact that an intuitive but wrong answer immediately comes to mind; and one has to engage in deliberation in order to override this intuitive and wrong answer and find out the correct one. We changed the original CRT, while preserving the fundamental structure, in order to avoid participants giving the right answer only because they are familiar with the test. The number of correct solutions is taken as a measure of deliberation. Previous work found that scores in the CRT correlate with scores in the Need for Cognition test and therefore can be considered a measure of reflective thinking (Pennycook et al., 2016).

After the CRT, participants answered a standard demographic questionnaire, at the end of which they received the completion code needed to submit the survey and receive their bonus. The bonus for the dictator game decision was computed as follows. Remember that all participants in this study played in the role of Participant B. Each Participant B was then paired with a Participant A who played in Capraro & Vanzo (2019), that was randomly selected among all participants who played in the frame chosen by Participant B. The bonus for the 11-20 task was computed by randomly pairing the participants in this experiment. All decisions involving money were incentivized.

---

[3] The frames are identical to those already used by Capraro and Vanzo (2019). These frames were chosen using the SentiWordNet lexical resource introduced by Baccianella et al. (2010). This lexical resource divides the English vocabulary in "synsets" (i.e., sets of synonyms) according to their sentiment polarity (Pang & Lee, 2004; Liu et al., 2007; Vanzo et al., 2014). We chose six synsets defined by words that describe well actions in an extreme dictator game: *steal*#1 = "take without the owner's consent"; *take*#8 = "take into one's possession"; *demand*#1 = "request urgently and forcefully"; *give*#3 = "transfer possession of something concrete or abstract to somebody"; *donate*#1 = "give to a charity or good cause"; and *boost*#2 = "be beneficial to". Moreover, these synsets are approximately symmetric in terms of their sentiment polarity: two of them (donate and boost) have non-zero positivity and zero negativity, two of them (steal and demand) have non-zero negativity and zero positivity, and two of them (take and give) are fully neutral, i.e., their positivity and negativity are both zero.

**Results**

*Participants*

We recruited N = 557 participants.[4] After excluding, as pre-registered, those who failed one or more comprehension questions and those with duplicate TurkIDs and IP addresses, we remain with N = 332 participants (193 males, 135 females, 4 prefer not to say; mean age = 39.35 (SD = 11.72); 173 participants in session 1 and 159 participants in session 2). An exclusion rate of 40% is in line with similar studies on AMT (Horton et al., 2010).

*Test of Hypothesis 1*

The distributions of decisions do not differ between the two sessions (Kolmogorov-Smirnov: combined D = 0.020, p = 1.000). We thus start by reporting the analysis of the two sessions together, and then we discuss the robustness of the results across sessions. Figure 1 reports the frequencies of participants choosing each frame. If participants were to choose uniformly at random, we would observe each frame chosen with a frequency of 1/6. The horizontal line in the figure reports this uniform frequency. The figure clearly shows that the observed distribution is different from the uniform distribution. Hypothesis 1 states that people choose the *steal* frame more frequently than one would expect from uniformly random choice. To test this hypothesis, we use a one-sample t-test to compare the observed frequency of each frame with the theoretical frequency assuming uniformly random choice, that is 1/6. In doing so, we find that the *steal* frame is chosen more often than one would expect from uniformly random choice (t = 4.78, p < 0.001). We note that Hypothesis 1 is confirmed also if we analyze the two sessions separately (session 1: t = 4.25, p < 0.001; session 2: t = 2.12, p = 0.012).

Although not pre-registered, we also look at the other frames. The *boost* and the *demand* frames are chosen less frequently than one would expect from uniformly random choice (boost: t = -4.35, p < 0.001; demand: t = -4.11, p < 0.001); the *donate*, the *take* and the *give* frames are chosen with a frequency that is not significantly different from the one expected from uniformly random choice (all p's > 0.09). All these results hold also if we analyze the sample by splitting session 1 and session 2, except for the *donate* frame, that appears to be chosen more frequently than one would expect from uniformly random choice in session 2 (t = 1.78, p = 0.038), but not in session 1 (t = 0.02, p = 0.491).

---

[4] Note that this number is higher than the pre-registered sample size (400 participants for the two sessions combined), because it also includes participants who get excluded during the survey because they fail one or more comprehension questions: these participants are not counted by Amazon Mechanical Turk, because they do not reach the end of the survey, but they are present in the datafile downloaded from Qualtrics. The pre-registered sample size corresponds to the number of participants as counted by Amazon Mechanical Turk, because this is the only number that we can actually control a priori.

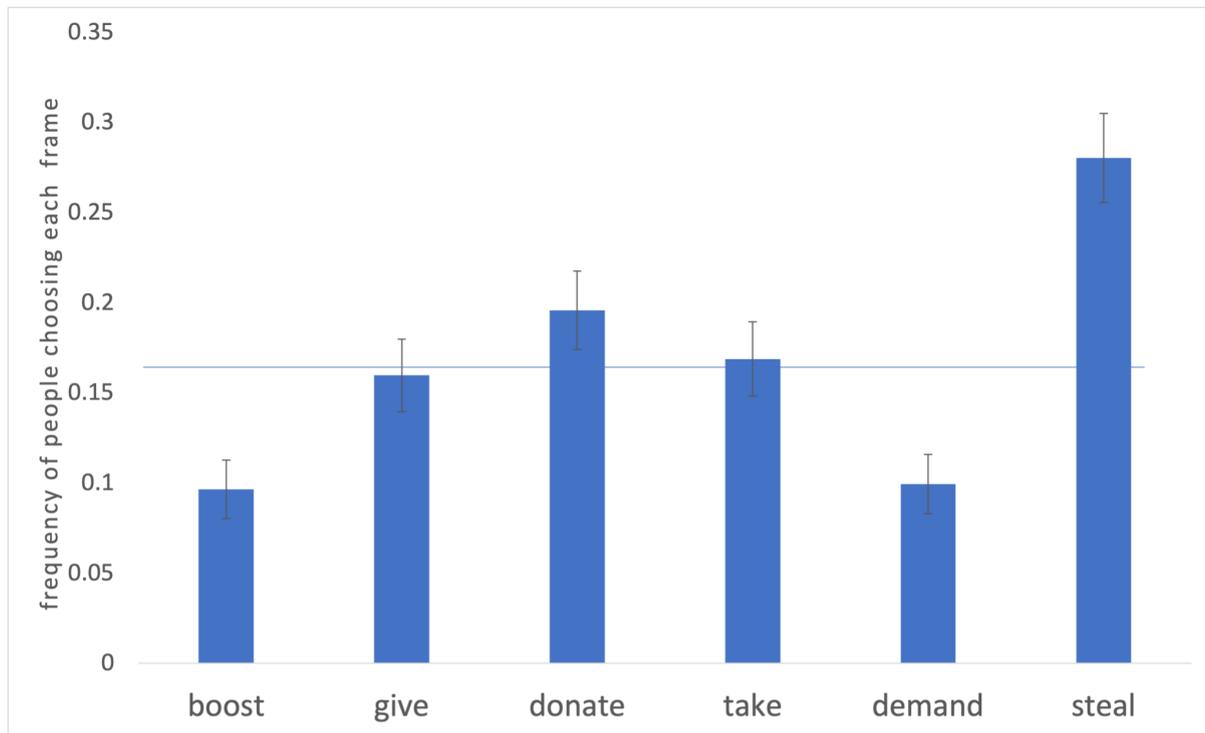

**Figure 1.** Frequencies of participants choosing each frame. The horizontal line, corresponding to 1/6, denotes the frequency of people who would choose any frame if they were to choose uniformly at random. Error bars stand for the standard errors of the means.

*Test of Hypothesis 2*

As a secondary pre-registered hypothesis, we had that participants with higher CRT are more likely to choose the *steal* frame. We pre-registered that we would test this hypothesis using a logistic regression predicting the "steal dummy", which takes value 1 if a participant chooses the *steal* frame and 0 otherwise. In doing so, we find that the hypothesis is not supported (b = 0.199, p = 0.304). The results are similar when we add controls on gender, age, and strategic sophistication (b = 0.221, p = 0.265). We only find an effect of age such that older people are less likely to choose the *steal* frame (b = -0.039, p = 0.002). One limitation of this analysis is that it collapses together all frames but the *steal* frame. To address this, we conducted a multinomial logistic regression predicting the frame decision as a function of CRT score, treating the *steal* frame as a reference frame. The results show that people who choose the *boost* frame tend to score lower to the CRT than people who choose the *steal* frame (b = -0.792, p = 0.015). All other p's > 0.1.[5]

**Study 2**

---
[5] These results were robust to including controls on sex, age, and strategic sophistication.

Study 1 shows that receivers are more likely to choose the *steal* frame than one would expect from random choice. This, however, does not imply that receivers choose the *steal* frame because they want to maximise their payoff. It is possible that they choose the *steal* frame because they prefer this frame for some other reasons. In the next two studies, we therefore provide two additional pieces of evidence in support of our interpretation. We show that receivers who choose the *steal* frame are aware of the fact that this frame is more likely to give them a higher payoff (Study 2) and that they would not choose the *steal* frame if they were to choose the frame as dictators (Study 3).

**Methods**

Study 2 is identical to Study 1, with two key differences: (i) after the decision, we also asked participants: "Now that you have chosen the wording of the options as they are presented to Participant A, which option do you think Participant A will implement?" Available answers were: "Option 1 (allocate the money to her/himself)" / "Option 2 (allocate the money to you)"; (ii) participants could choose only among the *steal* frame and the *boost* frame. We conduct only these frames in order to increase statistical power.[6] Specifically, we choose these frames for two reasons. First, from Study 1, we know that the *boost* frame is the least chosen frame. Second, from Capraro and Vanzo (2019), we know that the *boost* frame is also the frame that gives rise to the least amount of altruistic behavior. Therefore, choosing these frames allows us to increase sample size by frame and, at the same time, maintain the effect size as high as possible, and therefore increase statistical power. As in Study 1, we included five comprehension questions. Only participants who responded to all comprehension questions correctly were allowed to proceed with the experiment and make the actual decisions. After making their decisions, participants take the CRT and, finally, a standard demographic questionnaire. Compared to Study 1, we dropped the 11-20 task, since it was not correlated with the decision. After the demographic questionnaire, participants receive the completion code needed to claim for their bonus on AMT. As in Study 1, the bonuses were computed by matching the frame decisions of the current participants with the choices made in the corresponding frame collected by Capraro and Vanzo (2019). Our pre-registered hypothesis is that the participants who choose the steal frame expect, on average, a higher amount than participants who choose the *boost* frame. The design, the hypothesis, the analysis, and the sample size are pre-registered at: https://aspredicted.org/blind.php?x=ti728n.

**Results**

*Participants*

---

[6] We also conducted a pilot with all the frames. The results were in the expected direction, but, due to insufficient statistical power, in some cases we did not reach statistical significance. For example, the frequency of Participants B who expected altruistic behavior from Participant A, among those who had chosen the *boost* frame, was 0.32, compared to 0.62 among those who had chosen the *steal* frame. Despite this large numerical difference, the p-value was above 0.1.

We recruited N = 224 participants. After excluding, as pre-registered, those who failed one or more comprehension questions and those with duplicate TurkIDs and IP addresses, we remain with N = 162 participants (79 males, 81 females, 2 prefer not to say; mean age = 41.48 (SD = 12.46)).

*Test of the Hypothesis 3*

As hypothesized, we find that receivers who choose the *steal* frame are more likely than receivers who choose the *boost* frame to expect that dictators implement the altruistic option (50% vs 25%, Wilcoxon rank-sum: z = 3.23, p = 0.001). See Figure 2.[7]

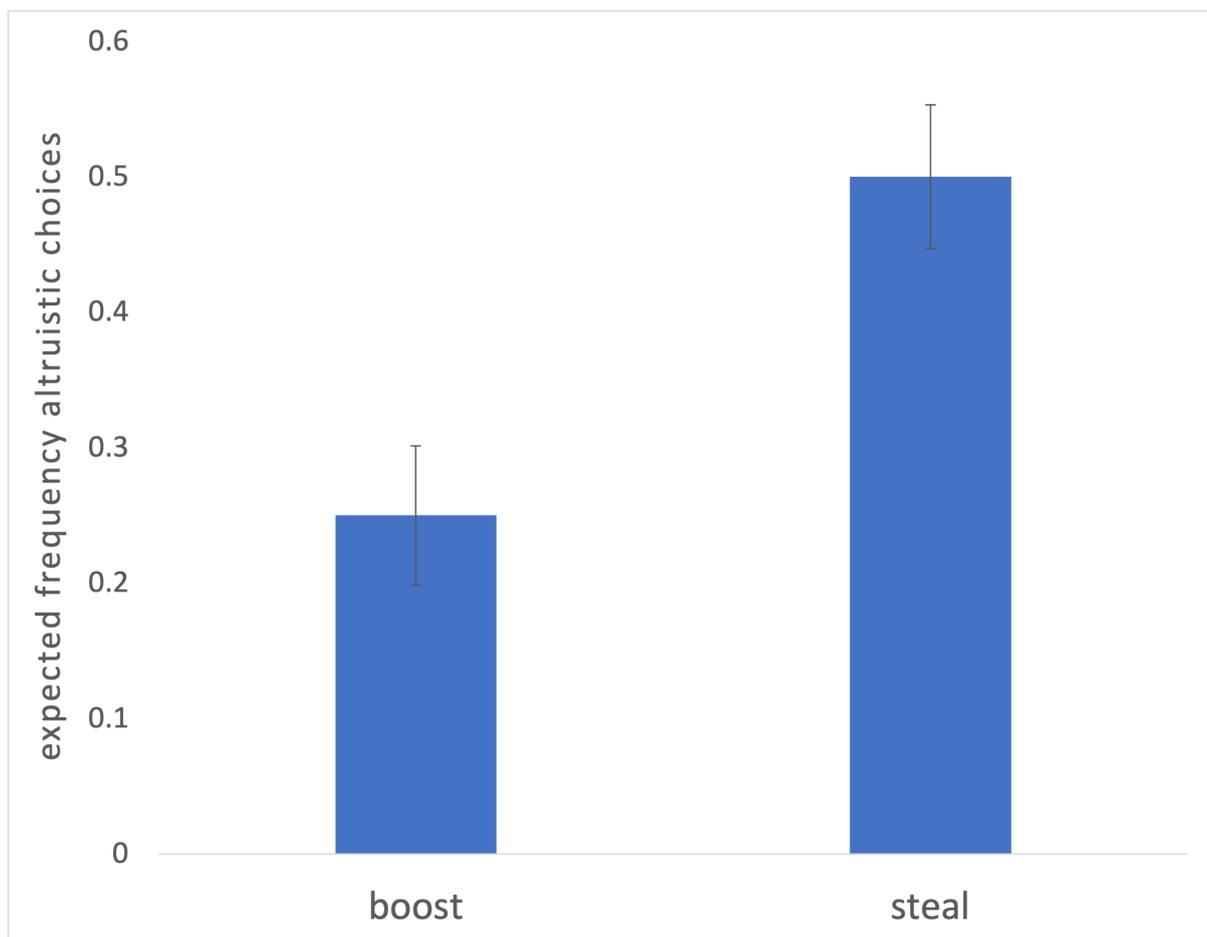

**Figure 2.** Frequencies of receivers believing that the dictator will be altruistic, by frame.

---

[7] Although not pre-registered, we also report that receivers tended to choose the *steal* frame more often than the *boost* frame (56% vs 44%, which correspond to 90 vs 72 participants), although the difference was not statistically significant (two-sample t-test, p = 0.079). The lack of statistical significance may be due to a combination of low statistical power and the use of two frames only, which may have reduced the room for statistical differences and may have added random noise due to forcing receivers to choose between the two most extreme frames. Also putting together Study 2 and Study 3, the trend was confirmed (54% vs 46%), but it still was not significant (p = 0.124).

# Study 3

Study 2 shows that receivers who choose the *steal* frame are more likely than receivers who choose the *boost* frame to expect a higher payoff. This, however, does not mean that these receivers choose the *steal* frame because they expect a higher payoff. It is possible that they choose the *steal* frame for some other, non-strategic, reason (e.g., because they like the word "steal"). To rule out this alternative, in Study 3 we compare the frame choice of receivers with that of dictators. If people choose the *steal* frame for non-strategic reasons, then we would observe that dictators are as likely as receivers to choose the *steal* frame. If, by contrast, the choice of the frame is strategic, we would observe that receivers are more likely than dictators to choose the *steal* frame.

## Methods

Participants are randomly divided between two conditions. The *receiver* condition is identical to Study 2, but without the measure of beliefs, that is, participants play as receivers in the dictator game and have to choose the description of the available options; as in Study 2, the only available options were the *boost* frame and the *steal* frame. In the *dictator* condition, participants have to choose the frame as dictators; again, the only available options were the *boost* and the *steal* frame. In order to not influence participants' decisions, dictators (receivers) are not informed that receivers (dictators) are also choosing the frame. After making their decision, participants take the CRT and, finally, a standard demographic questionnaire, at the end of which they receive the completion code needed to claim for their bonus. The bonuses were computed by matching dictators with receivers, according to the frames they chose; in case this procedure leads to a one-to-multiple matching, we implemented the average decision, which guarantees that the payment mechanism is incentive-compatible (Capraro & Barcelo, 2021). Our pre-registered hypothesis is that dictators are more likely than receivers to choose the *steal* frame. The design, the hypothesis, the analysis, and the sample size were pre-registered at: https://aspredicted.org/blind.php?x=yc6zu5.

## Results

### *Participants*

We recruited N = 225 participants. After excluding, as pre-registered, those who failed one or more comprehension questions and those with duplicate TurkIDs and IP addresses, we remain with N = 177 participants (87 males, 88 females, 1 prefers not to say, 1 did not respond; mean age = 40.58 (SD = 13.13)).

### *Test of the Hypothesis 4*

As hypothesized, the proportion of receivers who choose the *steal* frame is significantly higher than the proportion of dictators who choose the *steal* frame (50% vs 30.5%; Wilcoxon rank-sum: z = 2.575, p = 0.010).

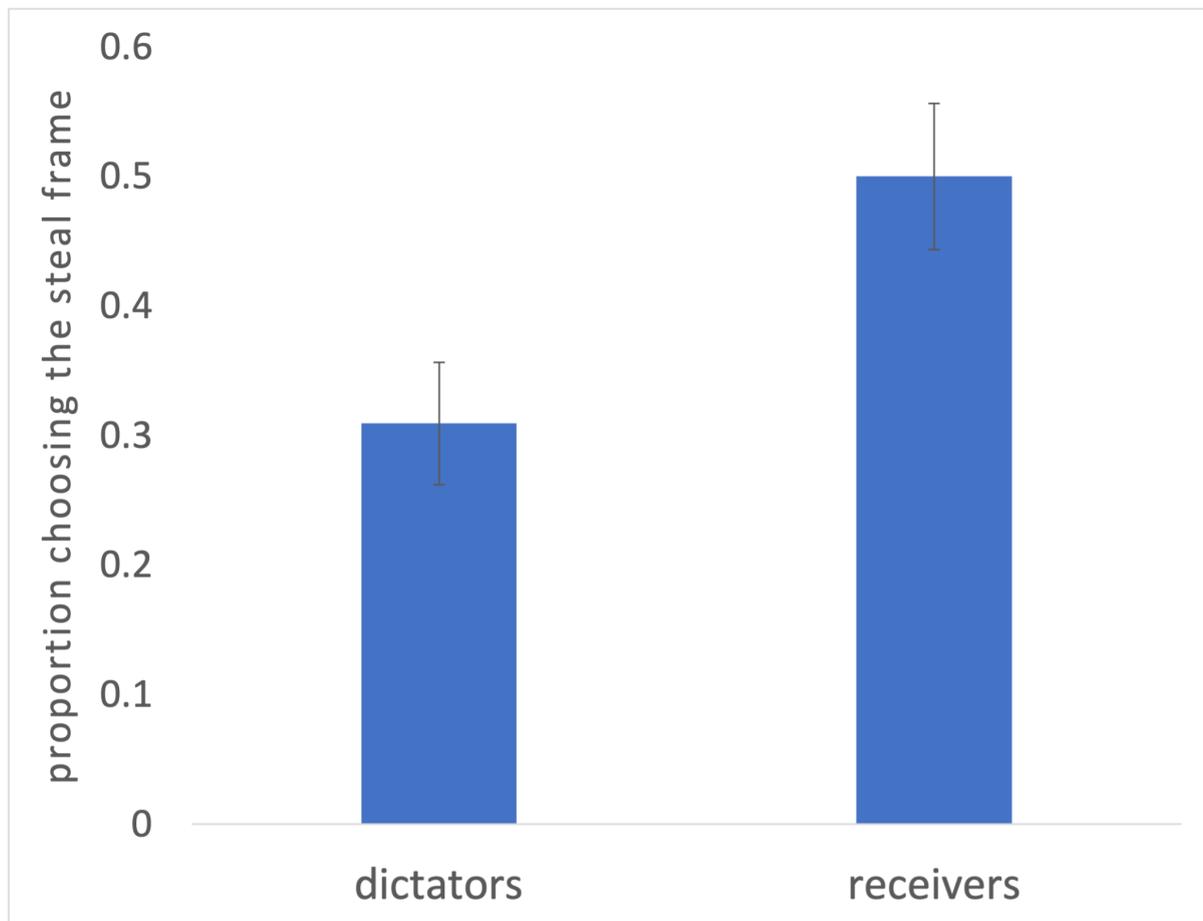

**Figure 3.** Proportions of dictators and receivers choosing the *steal* frame.

**Additional analysis**

As an additional, non pre-registered analysis, we also check the effect of deliberative thinking, gender, and age, on receivers' decision to choose the *steal* frame vs the *boost* frame, in Studies 2-3. We pool together the data of these studies but, as a robustness check, we control for the study. In line with Hypothesis 2 and with Study 1, we find that deliberative thinking is associated with the probability of choosing the *steal* frame over the *boost* frame (logit regression: coeff = 0.234, p =0.039; probit regression: coeff = 0.146, p = 0.038). This result holds also when adding controls on gender, age, and study (logit: coeff = 0.244, p = 0.038; probit: coeff = 0.151, p = 0.038). Moreover, in line with Study 1, we do find an age effect: younger receivers are significantly more likely to choose the *steal* frame (all p's < 0.02).

**Discussion**

We have reported three pre-registered experiments testing whether dictator game recipients, when given the opportunity to choose the words used to describe the dictator game to dictators, tend to choose words that are more likely to benefit themselves. Study 1 showed that receivers tend to choose words that, in an earlier experiment, led to higher donations. Study 2 found that receivers who choose these words indeed expect a higher payoff. Study 3 found that dictators and receivers choose different words, suggesting a strategic reason behind the choice of the words.

This contribution has practical and theoretical implications. From a practical perspective, there are many real-life situations in which a person has the opportunity to decide which words to use to present a decision problem to a decision-maker. In this case, this person may exploit this opportunity for their benefit. We captured the core of this situation by means of an "extreme" dictator game in which recipients have the opportunity to choose the words used to present the game to dictators, without the dictators being informed of this. Our results show that, at least in this idealized situation, some people indeed choose words that benefit themselves. Study 3 also allows us to estimate the proportion of people choosing the frame strategically, by looking at the difference between the percentage of receivers who choose the *steal* frame and the percentage of dictators who choose the *steal* frame[8], which appears to be relatively small, 20%. This suggests that choosing the frame strategically in order to maximize one's payoff may not be the typical strategy that people apply in real life.

From a theoretical perspective, our results cannot be explained by standard theories of social preferences, which assume that the utility of an action is a function of the monetary outcomes associated with that action (Fehr & Schmidt, 1999; Bolton & Ockenfels, 2000; Engelmann & Strobel, 2004): since all frames considered in this paper are identical from the perspective of monetary outcomes, social preference models predict that dictator game recipients are indifferent between the frames. This however contrasts with what we found. Our results provide evidence that some recipients are aware of the fact that dictators' actions depend on the words used to describe the available actions, and strategically choose the description that is more likely to maximize their monetary payoff. Therefore, our results call for new models of preferences that take into account also the words used to describe the available actions (Bjorndhal, Halpern, & Pass, 2013; Capraro & Perc, 2021). One potential explanation for our results is that people are aware that morally loaded words may activate self-image, moral preferences for doing what one thinks to be the right thing (Capraro & Rand, 2018; Capraro & Vanzo, 2019; Chang et al. 2019; Eriksson et al. 2017; Mieth et al. 2021; Tappin & Capraro, 2018). If people are aware of this "power of moral words", then they may choose words that are aligned with their social value orientation: for example, selfish receivers may choose words that induce dictators to be altruistic, while selfish dictators may choose words that make their selfish behavior justifiable. This could be formalized by assuming that the instructions of a decision problem carry the moral value of the available actions and of the context in which the decision is being made, and that the utility function of decision makers somehow depends also on the instructions, via their moral value. However, a mechanistic model linking quantitatively language to decisions is still missing (Capraro & Perc, 2021).

---

[8] Note that this difference contains two potentially distinct classes of people: those who, as receivers, choose the *steal* frame in order to affect the dictator's choice; and those who, as dictators, choose the *boost* frame in order to justify their future selfish behavior in the dictator game. (Indeed, we do find a trend suggesting that dictators who choose the *boost* frame may be less likely to implement the altruistic allocation than those who choose the *steal* frame; 12% vs 23%, p = 0.158.) Although both these classes of people choose the frame that favors themselves the most, they need not be the same class of people.

The second contribution of our work regards the individual characteristics of participants who are more likely to choose frames that maximize their monetary payoff. Our results provide some evidence of an effect of deliberative thinking. This may indicate that deliberation makes people more likely to realize that they can use words to affect dictators' decisions. Previous research shows that deliberative thinking does not seem to be associated with altruistic behavior in dictator games (Rand et al. 2016; Fromell et al. 2020). Our results suggest that, some of these deliberative people (likely those who are not altruistic) decide to exploit the opportunity and choose the *steal* frame, while others decide not to exploit the opportunity and choose one of the other frames. Indeed, it is important to note that not all people who realize that words can be used to affect dictators' decisions actually choose the *steal* frame: if it had been so, then, in Study 1, we would have observed a peak of decisions in the *steal* frame, while all other frames would have been chosen with equal frequency. This is not what Study 1 suggests: the *boost* and the *demand* frames are chosen far less frequently than the other frames. Our results also provide evidence of an effect of age, such that younger people are more likely to choose the *steal* frame. However, it is possible that this is simply due to the fact that younger people tend to be more selfish in the dictator game (Engel, 2010).

The fact that dictator game recipients are most likely to choose the *steal* frame and least likely to choose the *boost* frame in Study 1 perfectly mirror the results of Capraro and Vanzo (2019), who reported that dictators assigned to the *steal* frame are the most likely to implement the other-regarding decision, whereas those assigned to the *boost* frame are the least likely. However, there are also some mismatches between the results of Study 1 and those reported by Capraro and Vanzo. The clearest mismatch regards the *demand* frame. Capraro and Vanzo found that dictators assigned to the *demand* frame are far more likely to implement the other-regarding allocation than those assigned to the *boost* frame, and as likely to implement the other-regarding allocation as those assigned to the *donate*, the *give*, and the *take* frames. However, Study 1 shows that recipients are far less likely to choose the *demand* frame than the *give*, the *take*, and the *donate* frames, and they are equally likely to choose the *demand* and the *boost* frames. This mismatch might be due to a mismatch between beliefs and actual behavior, such that recipients believe that dictators assigned to the *demand* frame are less altruistic than they actually are. Another somewhat unexpected result is that receivers in Study 2 and 3 are almost equally likely to choose the *steal* frame as they are to choose the *boost* frame (54% vs 46%, two-sample t-test, p = 0.124). However, it is important to note that this does not mean that receivers are choosing the frame uniformly at random. The fact that receivers who choose the *steal* frame tend to be younger and more deliberative, besides believing they will receive more, than those who choose the *boost* frame, suggests that they do so for strategic reasons. The fact that, in contrast with Study 1, Study 2 and 3 find no a statistical difference in the proportion of receivers who choose the *boost* frame vs the *steal* frame might be due to a combination of factors, including low statistical power as well as the fact that the use of only two frames may decrease the room for statistical changes and increase random error, because receivers are forced to choose only between the extreme frames.

As any experimental work, also this has some limitations. First, in our experiments, dictators were not informed that recipients had the chance to choose the words used to present the game. It is likely that making the choice of the frame common knowledge would change the results. On the one hand, dictators who receive the *steal* frame might feel that they are being manipulated and this might lead them to retaliate negatively, and thus make them more self-regarding, instead of more other-regarding. On the other hand, recipients might anticipate this logic and this might bring them to avoid choosing the *steal* frames. Therefore, it is possible that what we found to be the most frequent choice in Study 1 (the *steal* frame) would become the least common one. At the same time, recipients may still try to affect dictators' behavior by choosing frames that they believe to be associated with more other-regarding allocations, but that are not perceived to be manipulative. In sum, in case the frame choice is common knowledge the behavior of recipients and dictators might be very nuanced and far less than obvious. For this reason, we believe this to be an interesting venue for future research. A second limitation regards the stakes. As most experiments conducted online, we also used small stakes. Although these stakes are proportioned to the amount of time that participants spend on the experiment, one may wonder whether these results extend to decisions involving higher stakes. Previous work suggests that dictator game decisions collected on AMT are comparable to those collected in the lab. For example, Brañas-Garza et al. (2018) reported a mega-analysis of more than 3,500 dictator game observations with small stakes on AMT and found an average giving (30.8%) very close to the one found by Engel (2010) in his meta-analysis of 616 laboratory experiments (28.3%). However, there is some evidence of a weak stake-effect, such that people may become less altruistic at very high stakes (Larney, Rotella, & Barclay, 2019). Moreover, it is likely that people also become more deliberative as stakes increase. Therefore, it is possible that our results change at very high stakes. Exploring this is another interesting direction for future work. A third limitation of our work is that we used an extreme dictator game rather than a standard dictator game. We made this design choice in order to be able to derive our main hypothesis (Hypothesis 1) starting from the results of Capraro and Vanzo (2019) and, ultimately, because in the case of an extreme dictator game it is easy to write instructions that, across conditions, differ only for just a few words. However, a downside of this design choice is that we cannot say whether our results extend to the standard dictator games, or to other economic games where language has been shown to play a role. We hope that future research can investigate the boundary conditions of the effect revealed in this paper.

In sum, this work demonstrates that some dictator game recipients, when given the opportunity to choose the words being used to introduce the dictator game, tend to choose words that make them more likely to receive a higher payoff. We found some evidence that deliberative thinking and young age are associated with this tendency.

**Appendix: Experimental Instructions**

*Main task (choose the frame of the dictator game)*

The game you will be playing involves two participants, Participant A and Participant B, and an amount of money (50c).

There are two available options:
- Option 1 allocates the whole amount to Participant A;
- Option 2 allocates the whole amount to Participant B.

The role of Participant B is to decide the words that will be used to present Option 1 and Option 2 to Participant A.
The role of Participant A is to decide between Option 1 and Option 2, as presented using the words decided by Participant B.

IMPORTANT: Participant A will not be aware that Participant B is choosing the descriptions of the options. Participant A will only play the game as described by Participant B, without ever knowing where these descriptions come from.

No deception is used. Participant A and Participant B will be paid according to Participant A's decision.

Here are some questions to ascertain that you understand the rules. Remember that you have to answer all of these questions correctly in order to get the completion code. If you fail any of them, the survey will automatically end. If this happens, please submit your HIT using your TurkId at the place of the completion code, so that we can still pay you the 50c participation fee. Note, however, that, in this case, you will not receive any additional money.

Which option maximises Participant A's gain? (Available answers: Option 1/Option2)

Which option maximises Participant B's gain? (Available answers: Option 1/Option2)

Is Participant A aware of the fact that Participant B is choosing the description of the options? (Available answers: Yes/No/Not specified)

What is the role of Participant A? (Available answers: Choose between Option 1 and Option 2/Choose the words used to describe Option 1 and Option 2/Neither of the two)

What is the role of Participant B? (Available answers: Choose between Option 1 and Option 2/Choose the words used to describe Option 1 and Option 2/Neither of the two)

(The next screen was shown only to participants who responded correctly to all the comprehension questions. Those who failed one or more comprehension questions were automatically excluded from the survey.)

You passed all the comprehension questions.

It is now time to make your decision.

You are Participant B. This means that your role is to decide the words that will be used to describe Option 1 and Option 2 to Participant A. Remember that Participant A will then choose between the descriptions of Option 1 and Option 2 that you have chosen, and that you will be paid according to Participant A's decision.

Remember that Participant A will not be informed that the descriptions of the two options come from a predetermined set of descriptions. Participant A will simply be shown the rules of the game with the labels "Option 1" and "Option 2" replaced by your descriptions.

Which wording of Option 1 and Option 2 would you choose?

Option 1 = Participant A steals all the money from Participant B
Option 2 = Participant A does not steal all the money from Participant B

Option 1 = Participant A takes all the money from Participant B
Option 2 = Participant A does not take all the money from Participant B

Option 1 = Participant A demands all the money from Participant B
Option 2 = Participant A does not demand all the money from Participant B

Option 1 = Participant A does not give all the money to Participant B
Option 2 = Participant A gives all the money from Participant B

Option 1 = Participant A does not donate all the money to Participant B
Option 2 = Participant A donates all the money from Participant B

Option 1 = Participant A does not boost Participant B with all the money
Option 2 = Participant A boost Participant B with all the money

(Participants could tick on their preferred pair of choices. These options were displayed in random order.)

*11-20 task*

You are randomly matched to play a game against another participant.
In the game, both of you request an amount of money between 11 and 20 cents.

Each participant will receive the amount s/he requests, plus an additional 20 cents if s/he asks for exactly one cent less than the other participant.

What amount of money do you request? (Available answers: 11 cents/12 cents/13 cents/14 cents/15 cents/16 cents/17 cents/18 cents/19 cents/20 cents)

*Cognitive Reflection Test*

A postcard and a pen cost 150 cents in total. The postcard costs 100 cents more than the pen. How many cents does the pen cost? (A text box was provided, where participants could write their answer.)

If it takes 3 nurses 3 minutes to measure the blood pressure of 3 patients, how long would it take 300 nurses to measure the blood pressure of 300 patients? (A text box was provided, where participants could write their answer.)

Sally is making some tea. Every hour, the concentration of the tea doubles. If it takes 8 hours for the tea to be ready, how many hours would it take for the tea to reach half of the final concentration? (A text box was provided, where participants could write their answer.)